\documentstyle[prl,aps,floats,psfig]{revtex}  \long\def\levfig#1{#1}

\def\tp{\hbox{$t_{\perp}$} }         %
\def\ww{\hbox{${\cal E}_{\para}$} }          %
\def\td{\hbox{${\cal T}_{\perp}$} }          %
\def\cp{\hbox{$c_{\perp}$} }         %
\def\tpal{\hbox{$t_{eff}$} }         %
\def\teff{\tpal}

\def\FF{{\cal F}}
\def\G{{\cal G }}

\def\para{\parallel }

\def\resc{{\sqrt{D}} }     %

\def\reg{\downarrow}
\def\Reg{\Downarrow}

\def\kom{\v k}
\def\xt{\v x}  %
\def\yt{\v y}  %

\def\om{\omega}
\def\iom{i \omega}
\def\kx{{k_{\para}}}

\def\v#1{{\bf #1}}

\def\al{\alpha}

\def\eps{\varepsilon}

\def\andword{and }
\def\submittedto{submitted to }
\def\toappearin{to appear in }

\def\percent{\%
}

\long\def\taglia#1{#1}

\long\def\tagliasi#1{}

\def\beq{\begin{equation}}
\def\eeq{\end{equation}}
\def\beqn{\begin{eqnarray}}
\def\eeqn{\end{eqnarray}}

\def\eqref#1{ %
 (\ref{#1})}

\def\andword{and }
\def\toappearin{to appear in }

\long\def\singlecol#1{
\twocolumn[\hsize\textwidth\columnwidth\hsize\csname @twocolumnfalse\endcsname
              #1]}

\ifpreprintsty %
\long\def\singlecol#1{#1}
\fi %

\long\def\taglia#1{}

\begin{document}  
\draft   %

\title{ 
 Crossover from Luttinger-  to Fermi-liquid behavior in strongly 
anisotropic systems \\
in large dimensions
}

\author{Enrico Arrigoni 
}

\address{ 
Institut f\"ur Theoretische Physik,
Universit\"at W\"urzburg,
D-97074 W\"urzburg, Germany \\
e-mail: arrigoni@physik.uni-wuerzburg.de 
}

\singlecol{
\date{ \it to appear in Phys. Rev. Lett., sched.  July 12 (99) } %
\maketitle
\begin{abstract}
We consider the low-energy region of an array of Luttinger liquids
coupled by a weak interchain hopping.  The leading logarithmic
divergences can be re-summed to all orders within a self-consistent
perturbative expansion in the hopping, in the large-dimension limit.  
The anomalous exponent scales to
zero below the one-particle crossover temperature.  
As a consequence, coherent
quasiparticles with finite weight appear along the whole Fermi
surface.  Extending the expansion self-consistently 
to all orders turns out to be crucial in order to restore the 
correct Fermi-liquid behavior.
\end{abstract}
}%

One-dimensional metals are 
characterized by the failure of the quasiparticle concept, 
signaled by the vanishing of the single-particle spectral 
weight at the  Fermi surface (FS). 
This produces a number of additional anomalous single- and
two-particle properties , as the absence of a step  
in the momentum distribution, spin-charge separation,   
the  vanishing of the local density of states $\rho(\om)$
as a power law like $\rho(\om)\propto \om^\al$,
as well as 
  nontrivial 
power-law behavior of several correlation functions. 
All these anomalous properties 
(except for spin-charge separation)
are controlled by the single exponent  
$\al$ .
Metals showing these properties are called Luttinger liquids (LL)\cite{luttinger}.
On the other hand, 
most isotropic two- and higher-dimensional metals show  Fermi-liquid
(FL) properties, 
their low-energy regime can be described by well-defined
  quasiparticles, and the exponents of 
the correlation functions 
coincide
with those inferred from dimensional
 analysis.  

We thus encounter two  distinct universality classes when going from one to higher 
dimensions, and  it's certainly 
a fascinating issue to study the crossover region between 
them.
Castellani, Di Castro and Metzner\cite{ca.dc.me} have approached this
problem by carrying out an  
analytic continuation from one dimension $D=1$ to $D=1+\eps$.  
They have shown that the 
FL regime is always recovered for  $D>1$, except for broken-symmetry states.

An alternative  way is to 
approach the crossover  
for the 
 {\it anisotropic} case by starting from a 
$D-1$ dimensional array
 of uncoupled LL
  and switching on a small single-particle hopping $\tp/\resc$
 between them \cite{resc}.
This also represents, in principle,  the experimental situation, for
example of 
the organic quasi-one-dimensional conductors.
The crucial question is: does the system go over to a 
 FL for arbitrarily small \tp
and sufficiently low temperatures?
One expects that, at least for temperatures $T$ greater than a certain
crossover temperature $T_{1D}$
the
system still shows LL properties. 
In fact, it has been shown (see, e. g. Ref.\cite{bo.bo.95}), that 
the region of
validity of the LL extends down to  $T_{1D}\approx \tpal \equiv 
t_{\perp}^{1/(1-\al)}$\cite{units,compl,ko.me.93}.
Therefore, the interesting, nontrivial region is the one for which 
the characteristic energy scale of the system, \ww\cite{units} is
smaller than \tpal. Only in this region, can a possible FL behavior occur.

Since \tp is small, the natural idea is to carry out a perturbative
calculation in \tp.
Many results have been obtained by restricting to the lowest order 
or to the first few orders in 
\tp \cite{wen.90,bo.bo.95,cl.st.94}. 
 An expansion in powers of \tp has been introduced in
 Ref. \cite{bo.bo.95} and 
systematized diagrammatically by
 the  author \cite{cross.97}.
This was done 
 by extending a method introduced by Metzner to expand around the 
atomic limit of the Hubbard model \cite{metz.91}.
The lowest order in \tp for the (inverse-) self-energy has been studied by 
Wen \cite{wen.90} and by Boies {\it et al.} \cite{bo.bo.95}. Within
their approximation (below referred to as LO), 
the quasiparticle structure is recovered 
for arbitrarily small \tp 
 on most of the FS.
However, 
the anomalous exponent $\al$  still dominates  the 
 low-energy ($\kom/\teff\ll1$)\cite{units} behavior 
for small $\cp$
since the self-energy behaves as $|\kom|^{1-\al}$, i. e., its {\it
  asymptotic} behavior is still LL-like.
As a physical consequence, 
close to  the 
regions  for which \cp vanishes, 
the spectral function is incoherent and the quasiparticles are thus
poorly defined (see Fig. 2). In particular,
 the quasiparticle 
 weight (or residue) $Z$ vanishes as $Z\propto (\tp \cp)^{\al/(1-\al)}$. 
However,  the LO approximation
is not sufficient to make definite conclusions {\it not even for small \tp}.  The 
reason  is that the perturbation \tp is relevant, i. e.,  each
order in \tp 
carries a strongly divergent power of 
the energy $\ww^{\al-1}$.
Any  perturbative expansion restricted  to
lowest order  is thus uncontrolled
at low energies $\ww\ll\tpal$.
This is the reason why 
theoretical results
 are still contradictory 
about the nature of the ground state
in this energy region.
Since, as discussed above, this is precisely the 
relevant region for a possible FL behavior, it is worthwhile
investigating it in a controlled way.

In this Letter, we analyze for the first time 
unambiguously
the region $\ww\ll\tpal$
by extending the calculation of the  self- 
energy to infinite order in  \tp. 
Specifically, we sum self-consistently the infinite series of loop diagrams of Fig 1a. The sum of these diagrams with fully dressed hopping (cf. Fig. 1b) gives the 
exact result in the (anisotropic)
 large-D  limit.  
Drastic changes are obtained with respect to the LO result,
which
 corresponds
to taking just the 
first term on the r.h.s. in Fig. 1a
 \cite{wen.90,cross.97}.
Our main result (Fig. 2) is that the anomalous exponent $\al$ scales to zero 
and the asymptotic behavior of the correlation functions is dominated by
mean-field-like  exponents and thus it becomes FL like.
In particular, our result restores 
coherent quasiparticles with a
  finite   weight $Z$ around
  the regions where $\cp=0$.
This striking outcome points out the importance of extending the expansion 
self-consistently 
to all order in the cumulants. 
Stopping  the series at any finite order  would still get an anomalous
behavior and  a vanishing 
$Z$.

\levfig{
\begin{figure}[htb] 
\par \vspace*{-.2cm} \par
   \centerline{\psfig{file=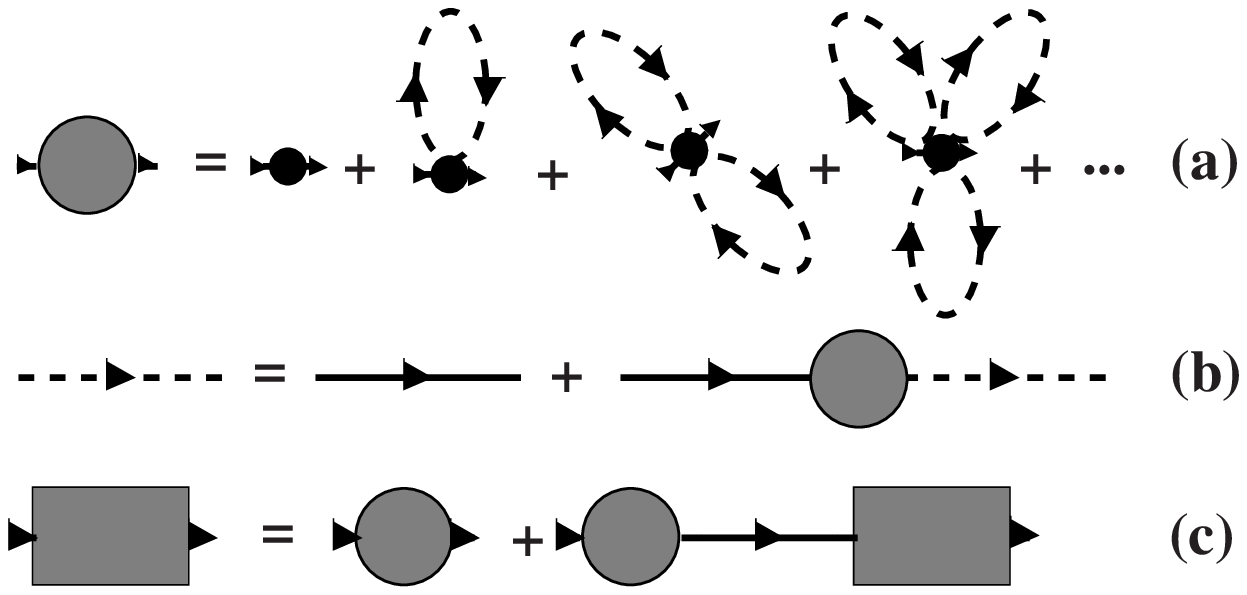,width=8.4cm}}
{\small
FIG.1. Diagrams contributing to the inverse-self-energy $\Gamma(\kom)$ (gray circle)
 in the
       $D=\infty$ limit within an expansion in the dressed hopping \td
       (a). 
(b)  Dressed hopping (dashed line) and its diagrammatic expression in terms of the bare hopping \tp (full line).
 (c) Dyson equation for the full Green's function $\G(\kom)$
(cf.       Ref. \protect{\cite{cross.97}}).
}
\par \vspace*{-.1cm} \par
 \end{figure}
}

We consider an hypercubic array of  LL
with the total dimension  $D$,  characterized by an
exponent $\al$, and with 
 a  hopping $\tp/\resc$ coupling them\cite{resc}.
 Here, we take the spinless case 
 since it allows for crucial simplifications in the calculation.
Since we are interested in the effects and in the 
fate of the anomalous exponent $\al$, 
we believe that 
spin-charge separation should not play a role. 

Since it is not possible to sum 
all diagrams in the expansion, we want to select a
workable subset of diagrams according to some {\it physical} limit in order to
avoid an arbitrary choice.
  The $D\to\infty$ limit allows us to
restrict ourselves to
the diagrams 
indicated
in Fig. 1a. 
Our $D\to\infty$ procedure is different from the standard dynamical
mean-field theory \cite{dinf} in several aspects.
First, and most important,
 in our case the system is strongly anisotropic, since
the hopping in one (in the $\para$) direction is not rescaled by the
$\resc$ factor and is  much larger than in
the other $D-1$ ($\perp$) directions. Our system thus represents a
1-D chain embedded in an effective self-consistent medium. 
Since this captures the one-dimensionality 
and the anomalous exponent
of the initial system, it is
particularly suited to study the problem of the crossover from one to
higher dimensions.
Notice that 
the system  has more degrees of freedom than the standard
$D\to\infty$ single-impurity problem.
As a consequence, the self-energy is local with respect to
the $\perp$ coordinates 
but has a nontrivial dependence on 
the $\para$ ones\cite{units}.
The second difference is merely procedural: instead of solving the
dynamical mean-field equations, we sum
 {\it all} the diagrams
contributing to the
self-energy in
$D\to\infty$. The internal lines of these
 diagrams
 must be evaluated self-consistently (see Fig. 1).
Third, since we are interested in the asymptotic behavior, we  
only keep the 
 leading  singularities in  each  diagrams, as we will discuss below.

As anticipated,
 each power of \tp carries
a  divergent power  of the energy $\propto \ww^{\al-1}$ \cite{bo.bo.95,cross.97}.
A crucial point in our approach is to consider an expansion in the {\it dressed}
hopping 
\td instead
of the bare one \tp (see below and Fig. 1). 
Since the scaling behavior of \td turns out to be $\td\propto 
\ww^{1-\al}$, this cancels exactly the  power-law divergences
met in the \tp expansion. 
However,  logarithmic divergences,  arising from the integration of the
cumulants still occur, as we shall see below.

Further ingredients of the perturbation theory are  the {\it bare}
(i. e., $\tp=0$) 
$\perp$-local $m+1$-particle cumulants 
of the   LL\cite{cross.97}
$\G^0_c(\xt_0 \cdots \xt_m | \xt_{0'} \cdots \xt_{m'})$. 
For these, we take the form valid for $T=0$ and large 
distances\cite{details}.
It will prove convenient to work in real (i. e.,
$\xt$) space, in
contrast to usual perturbation theory.
As discussed above, in 
 the $D\to\infty$ limit, 
the inverse-self-energy
$\Gamma(\xt_0)$ 
is $\perp$-local, and 
is obtained as the sum
of the loop diagrams 
in Fig. 1a as
\beqn
\label{gamma}
&& \Gamma(\xt_0) = 
\G^0_c(\xt_0|0) + \sum_{m=1}^{\infty}
(-1)^m 
\int_{1\reg m} 
\Bigl[ \prod_{k=1}^m
d^2 \yt_k 
\\ && \nonumber \times
d^2 \xt_k \ 
\td(-\xt_k,0) \Bigr] 
\G^0_c(\yt_0+\xt_0, \cdots, \yt_m+\xt_m | \yt_{0}, \cdots, \yt_{m}) \;,
\eeqn
where  ``$1\reg m$'' indicates  the region
$|\xt_1| > |\xt_2| > \cdots > |\xt_m|$ to which the integral can be restricted by 
exploiting
the symmetry under exchange of the coordinates $1,\cdots,m$.
In Eq. \eqref{gamma}, $\td(\xt,x_{\perp}=0)$
is the dressed  hopping written in real space.
As anticipated,
we are interested in the dominant
low-energy 
behavior
(here, 
$|\xt_0| \tpal \gg 1$)
of  correlation functions and thus we can restrict to
the leading logarithmic divergences in  the loop integrals\eqref{gamma} 
\cite{parquet}.
These 
 can be shown to come from further restricting to
the region $|\xt_p| < |\xt_0|$ for each
 $p$, and 
$|\xt_p| < |\yt_q+\epsilon_1 \xt_q -\yt_{q'} - \epsilon_2 \xt_{q'}|$ 
for each $p \geq q,q' $, $q\not=q'$, and $\epsilon_i=0,1$, which
will be referred to as ``$0\Reg m$''.
For convenience, we introduce the ``restricted renormalized cumulants'' (RRC) 
in this region defined as
\beqn
\label{gc}
&&
\G_c(\yt_0+\xt_0, \cdots, \yt_m+\xt_m | \yt_{0}, \cdots, \yt_{m})
\\ \nonumber &&
\equiv
\G^0_c(\yt_0+\xt_0, \cdots, \yt_m+\xt_m | \yt_{0}, \cdots, \yt_{m})
\\ \nonumber && 
-
\int_{0\Reg m+1}
d^2 \xt_{m+1} 
d^2 \yt_{m+1} 
\td(-\xt_{m+1},0) 
\\ \nonumber && \times
\G_c(\yt_0+\xt_0, \cdots, \yt_{m+1}+\xt_{m+1} | \yt_{0}, \cdots,
\yt_{m+1}) \;.
\eeqn
Comparing \eqref{gamma} and \eqref{gc}, 
one can verify that 
$\Gamma(\xt_0) = \G_c(\xt_0|0)$.

Within the leading-logarithmic approach discussed above, one can show that
 the RRC can be written in terms of the bare cumulants as
\beqn
\label{gcff}
&&
 \G_c(\yt_0+\xt_0, \cdots, \yt_m+\xt_m | \yt_{0}, \cdots, \yt_{m})
\\ && \nonumber
=
 \FF_m(l_{0},\cdots,l_{m}) 
 \G^0_c(\yt_0+\xt_0, \cdots, \yt_m+\xt_m | \yt_{0}, \cdots, \yt_{m})
\eeqn
where the renormalization factors $\FF_m$ are functions of the
relative coordinates $\xt_i$ through
 $l_{i}\equiv \al \log(|\xt_i| \tpal)$ \cite{details}. 
We can thus first carry out
the integration over the ``center-of-mass'' coordinate $\yt_{m+1}$  in
 \eqref{gc} by simply considering the effect on the 
the bare cumulant.
The leading logarithmic contributions to this integral
can be evaluated in a lengthy but straightforward way\cite{details}.

We now need to evaluate 
 the dressed hopping \td to insert in \eqref{gc}. This
 can be obtained in
 $\kom$ space in
 terms of  $\Gamma(\kom)$\cite{cross.97} 
  by solving the Dyson-like equation of  Fig. 1b\cite{units}
\beq
\label{tdgam}
\td(\kom,\cp)=  \tp \cp /(1-\Gamma(\kom)\tp \cp) \;.
\eeq
Within our self-consistent calculation\cite{selfcons} one must use
the fully dressed inverse-self-energy \eqref{gamma} in \eqref{tdgam}.
We are interested in 
the leading contribution  to 
$\td(\xt,0)$ at large $|\xt|$. This comes from two 
types of singularities in momentum space:
 (i) from the shifted poles of \eqref{tdgam} at 
$\cp\not=0,|\kom|\not=0$ and (ii) from the $\cp=0,|\kom|=0$ power-law
singularity of $\Gamma(\kom)$. 
Taking, to start with, the LO form for $\Gamma$, 
namely $\Gamma(\kom)\approx\G^0_c(\kom)\sim |\kom|^{\al-1}$,   
 the shifted-pole contribution 
(i) turns out to behave like $|\xt|^{2\al-4}$, while (ii) yields the strongest
contribution $\propto |\xt|^{\al-3}$.
This is due to the fact that the shifted-pole singularity (i) is
smeared 
out after
integration over the $\perp$ momenta.

We now evaluate \td in terms of the 
inverse self-energy $\Gamma$, so that it can be inserted in \eqref{gc}.
From \eqref{gcff}
with $m=0$, 
$\Gamma(\xt_0) = \G^0_c(\xt_0|0) \FF_0(l_0)$. 
By taking the FT of this expression into momentum space, inserting in
\eqref{tdgam},
and transforming $\td$ again into real space, we obtain,
within our leading-logarithm scheme \cite{units},
\beq
\label{td}
\td(\xt_i,0) = \frac{i \al}{\pi} |\xt_i|^{\al-3} e^{i r \arg\xt_i}
\left[\bar \FF_0(l_i) + \bar \FF_0'(l_i) \right] \;,
\eeq
where 
$\bar \FF_0(l_i)= 1/\FF_0(l_i)$, and $\bar \FF_0'(l_i)= \frac{d}{dl_i}  \bar \FF_0(l_i)$.
Thanks to the $|\xt|^{\al-3}$ term in \eqref{td},
the FT
of $\td$ now scales 
as $|\kom|^{1-\al}$, as already anticipated. Therefore, 
the diagrams written in terms of  the dresses propagator \td
all scale
in the same way for low energies and thus only logarithmic 
singularities are left \cite{actually}. 

Using Eqs.\eqref{td},\eqref{gc}, and \eqref{gc}, and integrating
 over $\yt_{m+1}$ and $\xt_{m+1}$,
yields a recursive self-consistent equation\cite{details} 
for
the renormalization functions $\FF_m$, which
we have solved 
 together with the self-consistency
condition
$\bar \FF_0(l)=1/\FF_0(l)$,  
by expanding in powers of the variable
$l $ up to $40^{th}$ order 
 and analyzed the series by means of several
types of differential Pad\'e approximants.
The result of this  analysis yields
for the large-positive-$l$ behavior of the  renormalization function
$\FF_0(l) \sim  e^{c l}$, where the exponent $c$ turns out to be
equal to $1$ within a great degree of accuracy (about $0.01 \percent $).

Introducing this result in the expression for
the inverse-self-energy
(Eq. \eqref{gcff} with $m=0$)
yields
$\Gamma(\xt) \to \G^0_c(\xt|0) (|\xt|\ \tpal)^\al \to 1/|\xt|$,
i. e. the anomalous 
exponent $\al$ exactly cancels out!.
The same things happens 
in   momentum space. Here, one obtains
$\Gamma(\kom) = \G^0_c(\kom) \bar\FF_0(\al \log\ |\tpal/\kom|) 
\to (\iom-r \kx)^{-1}$
 (since
 $\G^0_c(\kom) \propto |\kom|^{\al}/(\iom- r \kx)$).
This can be interpreted as the LL exponent $\al$
renormalizing to zero for energies smaller than \tpal.

The 
Green's function of the  coupled  array of LL
is given by the
 Dyson-like equation\cite{cross.97}
$
\G(\kom) = \left(\Gamma(\kom)^{-1} - \tp\cp\right)^{-1} 
$ (cf. Fig. 1c).
The main important consequence of our result is that 
the low-energy
behavior of correlation functions, as,  say,
 the Green's function, is now FL  like.
Specifically, 
by an homogeneous rescaling of all energies
$\kom \to s \kom$, and $\cp \to s \cp$,  
$\G(\kom) \propto s^{-1}$ for $s\ll1$,
i. e.,  the anomalous dimension vanishes.
This result reflects onto the spectral properties in the region close to
$\cp=0$.
\levfig{
\begin{figure}[htb] 
\par \vspace*{-1.5cm} \par
   \centerline{\psfig{file=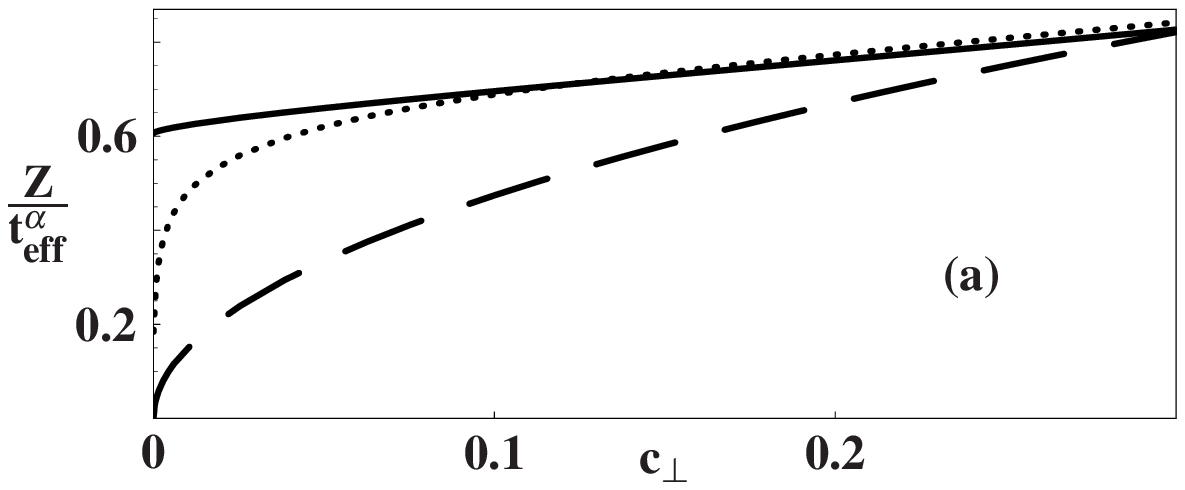,width=8.4cm}}
\par \vspace*{-2.3cm} \par
   \centerline{\hspace*{.3cm} \psfig{file=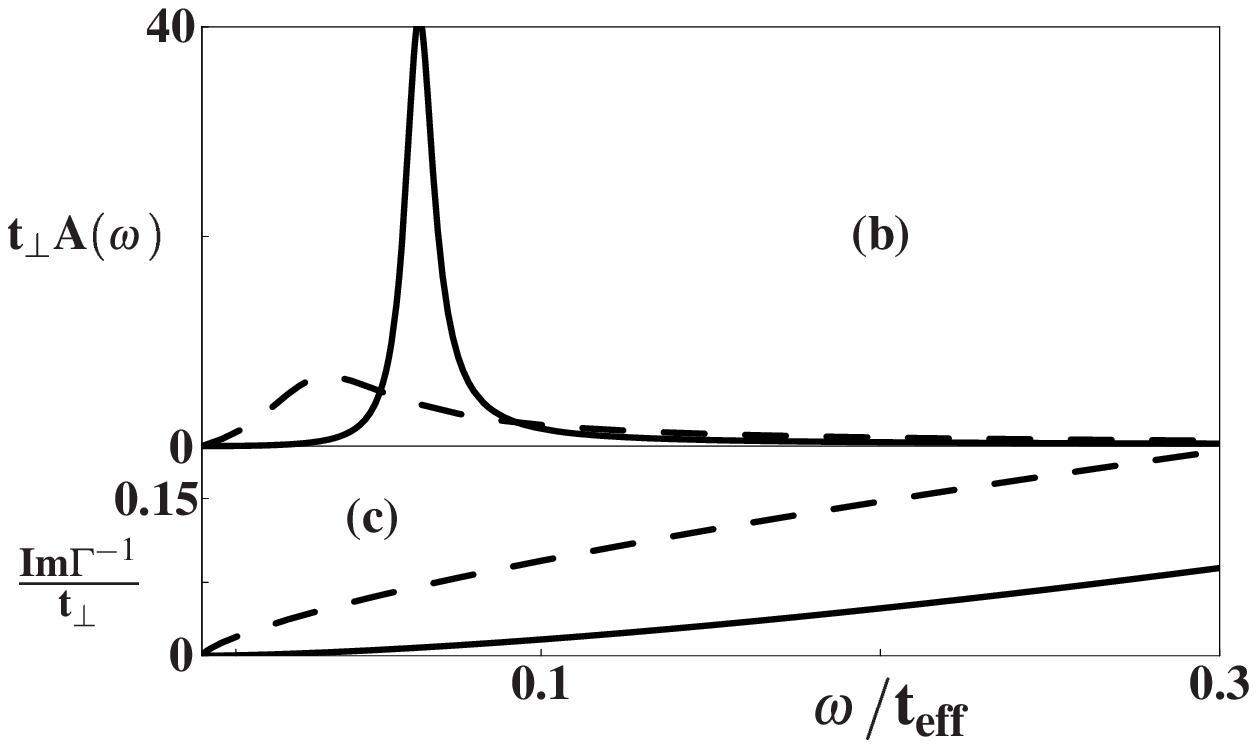,width=8.4cm}}
\par \vspace*{-.5cm} \par
{\small FIG. 2.
Quasiparticle weight Z (a) 
as a function of the off-chain kinetic energy \cp 
 (in units of \tp)  for
the LL exponent \mbox{$\al=1/3$}.
Spectral function $A(\omega)$ (b)
for $k_{\para}=0$
and $\cp=0.1$,
 and imaginary part of the
self-energy $Im \Gamma^{-1}$ (c) as a function of $\omega$. 
Our $D=\infty$ results (solid lines)
re compared with the LO result 
 (dashed). In (a) we also plot for comparison
  the result obtained including 
just the first two diagrams on the r.h.s. of
Fig. 1a 
 (dotted).
}
\par \vspace*{-.0cm} \par
 \end{figure}
}
The quasiparticle weight $Z(\cp)$ on the FS 
[$\kx_F(\cp)$] 
 is given by 
$Z(\cp)^{-1} =  \frac{d}{d i \om} 1/\G(\kx_F(\cp),\iom)_{\iom\to0}$, and is 
plotted in Fig. 2a 
for $\al=1/3$ and compared with two results obtained within 
the LO approximation (dashed line) and by taking in addition
 the first loop 
 diagrams  of Fig. 1a (dotted), respectively.
For small \cp, the LO result would give a
$Z$ vanishing as $Z(\cp) \propto \tpal^{\al} \cp^{\al/(1-\al)}$, 
 thus yielding poorly defined 
quasiparticles
around
$\cp=0$. Inclusion of the 
first loop
gives a vanishing Z too.
On the contrary,
 our result
yields a finite $Z$ for $\cp\to0$ as can also be seen from the figure.
This also reflects in the spectral function, 
plotted in Fig. 2b.
Our result shows a   coherent quasiparticle
peak in contrast to the rather broad and shallow maximum produced by the
LO approximation.
To understand this behavior, we have plotted the scattering rate 
($Im \Gamma^{-1}$) for $k_{\para}=0$ in Fig. 2c. As one can see, while for
the LO approximation $Im \Gamma^{-1}$ vanishes quite slowly
(as $w^{1-\al}$), in our result $Im \Gamma^{-1}$ vanishes faster than
linearly, as required for
well-defined quasiparticles.
However, 
 one should mention that our calculation, restricted to the leading
divergences, only 
yields reliable results for  $Im \Gamma^{-1}$ close to 
the FS points with  $\cp=0$.
In other FS points, we cannot reliably state whether 
$Im \Gamma^{-1}$ vanishes fast enough or not.
Arguments similar to the one 
of ordinary perturbation theory\cite{lutt.61}
cannot be extended to the present case,
due to the
momentum dependence of the vertices in the \tp expansion.

In summary, our result shows that 
 a quasi-one-dimensional metal with small perpendicular hopping \tp
crosses over from a LL behavior 
with anomalous exponent $\alpha$
for energies larger than
 $\tpal$  and flows to a FL fixed point with mean-field like
 exponents
for smaller energies in large dimensions.
This is due to the LL exponent $\al$ renormalizing to zero for
 $\ww\ll\tpal$.
As a consequence,
well defined quasiparticles are recovered along the whole FS. 
The result 
contrasts with  the LO approximation for which
 the anomalous exponent is still present in the asymptotic behavior of
 the self-energy below $\teff$ and the spectrum is incoherent for
 small \cp.

In principle, we cannot say
whether our result is valid also for $D=2$ or $D=3$. 
However, it turns out that 
 the non-$\perp$-local dressed hopping 
$\td(\xt,x_{\perp}\not=0)$  vanishes faster than
 the $\perp$-local one $\td(\xt,0)$  for large $|\xt|$.
Non $\perp$-local contributions are thus irrelevant and the present
result could be thus extended to finite dimensions.
This is  
valid for values of $\al$ smaller than a critical value $\al_{2p}$, 
 for which a transition to a two-particle regime  takes
 place\cite{cross.97}.
In addition, it may possibly exist  a critical dimension below which
non-$\perp$-local
diagrams may enhance the contribution from
the shifted $|\kom|,\cp\not =0$ poles,
 and give rise to nesting or
superconducting instabilities at selected regions of the FS. 
One additional important conclusion is 
that it is crucial to include
an infinite number of 
  diagrams in the \tp expansion, 
in order to obtain reliable results.

The author thanks W. Hanke for useful discussions.
Partial support by BMBF (05SB8WWA1) is acknowledged.

\ifx\undefined\andword \def\andword{and} \fi %
\ifx\undefined\submittedto \def\submittedto{submitted to } \fi %
\ifx\undefined\toapperarin \def\toappearin{to appear in } \fi %
\def\nonformale#1{#1}
\def\formale#1{}
\def\spa{} \def\spb{}
\def\spa{\ifpreprintsty\else\vspace*{-.5cm}\fi} %
\def\spb{\ifpreprintsty\else\vspace*{-1.6cm}\fi} %
\spa


\begin{thebibliography}{10}
\bibitem{resc}{
We rescale \tp by a factor $D^{-1/2}$, in order to obtain a nontrivial $D\to
  \infty$ limit, as pointed out in Ref. \cite{dinf}.}%
\bibitem{units}{
We give here our conventions: (i) The Fermi velocity $v_F$ and energy $E_F$ are
  set equal to $1$. (ii) The momenta of the right- and left-moving fermions in
  the isolated LL are measured from the respective Fermi momenta. (iii) $\ww$
  is the largest on-chain energy scale between the external frequency $\om$,
  the momentum $k_{\para}$ and the temperature $T$. (iv) The coordinates
  (position $x$ or $y$, momentum $k$, etc.) in the direction parallel to the LL
  have the subscript $\para$, while the ones in the ($D-1$-dimensional) plane
  perpendicular to it are labeled by $\perp$. (v) A boldface coordinate is a
  shorthand notation for: the coordinate in the $\para$ direction, the
  corresponding imaginary-time or frequency variable, and the index $r$
  indicating right- ($r=+1$) or left- ($r=-1$) moving electrons. For example,
  $\xt$ stands for $(x_{\para},\tau,r)$, and ${\kom}\equiv(k_{\para}, \om,r)$.
  Integrals over these coordinates implicitly include a sum over $r$. (vi)
  $\tp\cp \equiv 2 \tp \sum_{i=1}^{D-1} \cos(k_{\perp i})/\resc$ is the kinetic
  energy in the $\perp$ direction. (vii) Finally, we call $\perp$-local a
  quantity which is local in the $x_{\perp}$ coordinates or, equivalently,
  independent of the $k_{\perp}$ momenta.}%
\bibitem{compl}{
The region of LL behavior may extend down to quite low temperatures. For
  example, for typical parameters of the Bechgaard salts (see Ref. \cite{bech})
  as $\tp\sim 100 K$, $t_{\para}\sim 1000 K$, and, as it is believed $\al>1/2$,
  this region extends down to less than $10 K$.}%
\bibitem{selfcons}{
In contrast to the usual logarithmic parquet expansion \cite{parquet}, the
  Green's function is renormalized already at one loop in our calculation and
  thus its self-consistency has to be fully taken into account.}%
\bibitem{actually}{
Actually, as we will show below, the exponent $\al$ renormalizes to zero for
  $\ww\ll\tpal$, and thus $\td$ will eventually scale as $|\kom|$. However,
  since this happens for the renormalized cumulants\eqref{gcff} too, this
  conclusion remains unchanged.}%
\end{thebibliography}
\end{document}